\documentclass {iopart}
\usepackage{graphicx}

\begin{document}

\title{Magnetothermopower in Nd$_{1-x}$Eu$_{x}$NiO$_{3}$ compounds}

\author{V B Barbeta$^1$ , R F Jardim$^2$, M T Escote$^3$ and N R Dilley$^4$}

\address{$^1$ Departamento de F\'{i}sica, Centro Universit\'{a}rio da
FEI, S. B. Campo, SP 09850-900, Brazil}

\address{$^2$ Instituto de F\'{i}sica,
Universidade de S\~{a}o Paulo, CP 66318, S\~{a}o Paulo, SP 05315-970, Brazil}

\address{$^3$ Centro de Ci\^{e}ncias Naturais e Humanas, Universidade Federal do ABC, Santo Andr\'{e}, SP 09210-170, Brazil}

\address{$^4$ Quantum Design, Inc., San Diego, CA 92121, USA}

\ead{vbarbeta@fei.edu.br}

\begin{abstract}
We have measured magnetization $M(T,H)$, electrical resistivity $\rho(T,H)$, thermal conductivity $\kappa(T,H)$, and thermopower $S(T,H)$ of polycrystalline samples of Nd$_{1-x}$Eu$_{x}$NiO$_3$ ($0 \leq x \leq 0.5$) as a function of temperature and external magnetic field.  The data indicate a metal-insulator (MI) transition in a wide range of temperature ($200 < T_{\rm MI} < 325$ K).  The magnetic susceptibility $\chi(T)$ data, after the subtraction of the rare-earth contribution, exhibit a Curie-Weiss like behavior at temperatures above $T_{\rm MI}$.  Although a clear antiferromagnetic AF transition of the Ni sublattice is observed at $T_{\rm N} \leq T_{\rm MI}$, $\chi(T)$ still increases down to 5 K, suggesting a heterogeneous ground state.  The thermal conductivity of the NdNiO$_3$ compound is not affected by an external magnetic field of 90 kOe in a wide range of temperature, and its temperature dependence below 15 K is approximately quadratic, strongly suggesting the presence of disorder.  The data also show that $S(T)$ is negative above $T_{\rm MI}$ and varies linearly with temperature, as expected in a typical metal with negative charge carriers. Below $T_{\rm MI}$, $S(T)$ does not follow the expected behavior of insulating compounds.  There is a minimum close to 120 K, and $S(T)$ changes its sign at $T \sim 30$ K, indicating a competition between two types of charge carriers.  A pronounced peak in $S(T)$ at $T_{\rm S} \sim 20$ K does not follow the expected phonon-drag temperature dependence either above or below $T_{\rm S}$ and the peak remains unaltered under magnetic fields up to 90 kOe.  However, its magnitude is enhanced by $\sim 25$\% with applied magnetic field, exhibiting a clear magnetothermopower effect. The combined results indicate a coexistence of ordered and disordered phases below $T_{\rm N}$ and that an applied magnetic field is suitable for enhancing the thermoelectric properties of these perovskites close to $T_{\rm S}$.
\end{abstract}

\maketitle

Transition metal oxides \textit{R}NiO$_{3}$ (\emph{R} = rare earth, \textit{R} $\neq$ La) are interesting due to their transport and magnetic properties caused by strong electronic correlation effects.  These perovskites show a marked metal-insulator (MI) transition at a temperature $T_{\rm MI}$ which increases when the ionic radius of the rare earth decreases, ranging from $\sim 200$ K for \textit{R} = Nd to $\sim$ 500 K for \textit{R} = Eu.  They also show an antiferromagnetic (AF) phase transition at a temperature $T_{\rm N}$ related to the ordering of the Ni sublattice.  For \textit{R} = Nd, $T_{\rm N} \sim T_{\rm MI} \sim 200$K, and for \textit{R} = Eu, $T_{\rm N} \sim 205$ K, well below $T_{\rm MI}$ [1].
Neutron powder diffraction (NPD) experiments performed in \textit{R} = Nd showed an unexpected propagation vector $\textbf{k}$ = (1/2, 0, 1/2), probably related to an inhomogeneous ground state composed of alternating ferromagnetic and AF couplings between nearest-neighbor Ni$^{3+}$ along the three pseudocubic axes.  This unusual AF ordering has been attributed to the presence of an orbital ordering (OO) that would break down the inversion center at the Ni site in the orthorhombic symmetry [2].  Other NPD peaks were reported below $T \sim 20$ K and associated with a magnetic ordering of the Ni$^{3+}$ spins [3].

Results in smaller rare earth ions (\textit{R} = Ho, Y, Er, and Lu) also suggested a change in the crystal symmetry, from orthorhombic $Pbnm$ to monoclinic $P2_{1}/n$, across the $T_{\rm MI}$ transition, resulting in a charge disproportionation of Ni$^{3+}$ cations [4].  Electron diffraction and Raman scattering experiments indicated a symmetry break also for \textit{R} = Nd [5], and a direct observation of charge ordering (CO) in epitaxial NdNiO$_{3}$ films, using resonant x-ray scattering, has also been reported [6].  These results led to the proposition that charge disproportionation, and a consequent change in the crystal symmetry, may occur below $T_{\rm MI}$ for all the members of the \textit{R}NiO$_{3}$ family, making the picture of OO obsolete [6].  Recent soft x-ray resonant scattering experiments indicated that OO is absent at the MI transition, that the complex AF ordering previously proposed may not be correct, and that the magnetic structure is likely to be non-collinear [7].

Although some properties of these systems are well established, there are still many doubts regarding their low temperature ground state that seems to be inhomogeneous, dominated by the coexistence of at least two different phases. The data regarding the thermal properties as well as thermopower of these systems could be of interest to better understand their properties at low temperatures.  The lack of systematic measurements of these properties is the main motivation for the development of this work.

Polycrystalline samples of Nd$_{1-x}$Eu$_{x}$NiO$_{3}$ ($0 \leq x \leq 0.5$) were prepared by using sol-gel precursors, sintered at high temperatures ($\sim$1000 $^\circ$C), and under oxygen pressures up to 80 bar [8].  The samples were characterized by means of x-ray diffraction (XRD) measurements in a Brucker D8 Advanced Diffractometer using Cu $K_{\alpha}$ radiation ($\lambda = 1.540\,56$ \AA).  All samples were found to be single phase materials to XRD analysis [9].  Magnetization $M(T)$ measurements were taken in a commercial superconducting quantum interference device (SQUID) magnetometer.  The runs were performed cooling down and warming up the samples, with temperature ranging from $\sim 2$ to $\sim 400$ K and under dc magnetic fields as high as 10 kOe.  Electrical resistivity $\rho(T)$ measurements were performed by using the dc four-probe technique.  Thermopower $S(T)$ and thermal conductivity $\kappa(T)$ were measured by using a physical propertie measurement system (PPMS) equipment from Quantum Design.  Measurements of $S(T)$ and $\kappa(T)$ were also performed for the pristine compound (\textit{R} = Nd) in an applied magnetic field of 90 kOe.

The temperature $T_{\rm MI}$ was defined through the sharp peak observed in the curves of $-d(ln \rho) /dT$ vs $T$ in the warming up cycle.  For temperatures below $T \sim 50$ K, $-d(ln \rho) /dT$ changes and increases faster.  Increasing Eu content moves the MI transition to higher temperatures but preserves the behavior of $-d(ln \rho) /dT$ vs $T$ at low temperatures.  A pronounced thermal hysteresis is frequently observed close to the MI transition indicating the first-order character of the transition, as discussed elsewhere [9].

The $\chi(T)$ data of NdNiO$_{3}$ indicate that an extra contribution to the Ni sublattice must be considered below the magnetic ordering temperature $T_{\rm N}$.  Therefore, to study the NiO$_{6}$ sublattice is necessary to perform a precise subtraction of the contribution of the rare-earth ions from the measured $\chi(T)$ signal.  This kind of analysis has been done before, and indicated that the NiO$_{6}$ sublattice displays features of an unconventional antiferromagnet [9].  In fact, $\chi(T)$ still increases with decreasing temperature below $T_{\rm N}$, a behavior attributed to the existence of an inhomogeneous ground state comprised of alternating diamagnetic and paramagnetic sites [10].  Similar behavior of $\chi(T)$ has been observed in the Eu substituted samples studied here [9] and may have their counterpart in thermal properties, as described below.

The $\kappa(T)$ data, measured under the warming process for several samples, are displayed in Fig. 1.  A well defined jump in $\kappa(T)$ at high temperatures identifies the MI transition of the samples, that varies from $\sim$ 195 to 295 K for samples with $x = 0$ and 0.35, respectively.  In addition, samples with $x = 0.30$ and 0.35 exhibit a positive $d \kappa(T)/dT$ in the entire range of temperature whereas those with $x = 0$ and $x = 0.25$ display a negative $d \kappa(T)/dT$ in a small temperature range below $T_{\rm MI}$.  At temperatures close to $T_{\rm S} \sim 20$ K the $\kappa(T)$ data display a change in their curvatures from upward to downward, as inferred from a combination of the data shown in Fig. 1 and the inset.

\begin{figure} [htp]
\centering
\includegraphics [width=0.8\textwidth] {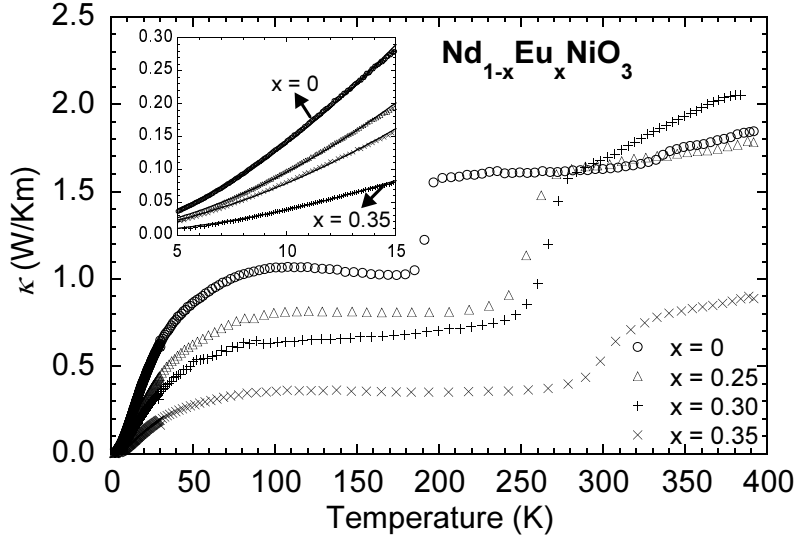}
\caption{\label{fig:epsart1} Temperature dependence of the thermal conductivity $\kappa(T)$ for selected samples.  The inset displays the low temperature data accompanied by the best fits obtained by using the relationship $\kappa(T) = aT^{b}$.}
\end{figure}

Considering a simple model where the system is comprised of non-interacting particles, $\kappa(T)$ can be expressed by the sum of two contributions: a phononic $\kappa_{l}$ and an electronic $\kappa_{\rm el}$ component.  At low temperatures,  $\kappa_{\rm el}$ can be neglected (insulating phase) and $\kappa(T) = \kappa_{\rm el}$ can be expressed as $\kappa(T) = Cvl/3$, where $C$ is the heat capacity, $v$ is the drift velocity, and $l$ is the mean free path of the thermal carriers [11].  Since $l$ and $v$ are expected to be temperature independent in the low temperature regime, $\kappa(T)$ would follow the $C(T)$ behavior, i. e.,  $\kappa(T) \sim T^{3}$.  Attempts to fit the low temperature $\kappa(T)$ data to a $T^{3}$ function have failed, even when an extra linear term was considered.  Nevertheless, the best fits were obtained by using $\kappa(T) = aT^{b}$, depicted in the inset of Fig. 1, with values of $a$ and $b$ shown in Table 1.  The factor $a$ follows a trend and decreases systematically with increasing Eu content, whereas $b$ was found to be $1.8 \pm0.1$.  The function is essentially quadratic and the fits start to deviate from the experimental data when the temperature is increased above $T \sim 15$ K.  A $T^{2}$ term for $\kappa(T)$ is expected when phonons are scattered by electrons.  However, this seems to be improbable here since these samples are insulating at low temperatures [8, 9].  Another possible origin for a $T^{2}$ term is due to the presence of itinerant magnetic excitations (magnons), although this contribution would be effective only at very low temperatures [11].

\begin{table}
\caption{\label{tab:table1}Parameters for the fit $\kappa(T) = aT^{b}$ at low temperatures in Nd$_{1-x}$Eu$_{x}$NiO$_{3}$ compounds.}
\begin{indented}
\lineup
\item[]\begin{tabular}{ccc}
\br
 $x$ & $a$ & $b$ \\
&(W/K$^{3}$m)&\\
\hline 0& 0.00221 & 1.80 \\
0.25 & 0.00143 & 1.83 \\
0.30& 0.00137 & 1.76 \\
0.35 & 0.00051 & 1.88 \\
\br
\end{tabular}
\end{indented}
\end{table}

The $\kappa(T)$ behavior strongly suggests the presence of disorder in the samples, since exponents similar to the ones found here are frequently observed in noncrystalline dielectric solids [12].  Under these circumstances, the low temperature $\kappa(T)$ would be mainly governed by phonon-defect scattering, being proportional to $T^{2}$.  The presence of disorder in these nickelates has been proposed before and may to be related to the variable range hopping (VRH) mechanism observed in $\rho(T)$ data at low temperatures [13].

Disorder can be an intrinsic property of the insulating phase of these compounds, as inferred from extended x-ray-absorption fine structure (EXAFS) in NdNiO$_{3}$.  It has been proposed that a CO modulation only occurs in a short range scale, but the system remains disordered in long range scale [14].  Within this scenario, the insulating state, below $T_{\rm MI}$, is better described as composed of insulating CO and Jahn-Teller distorted pockets.  The effect of lowering temperature below $T_{\rm MI}$ favors the growth of the insulating pockets.  In addition to this, the absolute values of $\kappa(T)$ are rather small at low temperatures ($\kappa \sim 0.1$ W/Km at $T \sim 10$ K), indicating that phonons are strongly suppressed.  Such a behavior seems to be consistent with an inhomogeneous ground state due to cooperative bond-length fluctuations [15].

Some of the features described above were also observed in thermopower $S(T)$ data, as shown in Fig. 2.  Above $T_{\rm MI}$, $S(T)$ is negative and varies linearly with temperature, indicating negative charge carriers and a metal-like behavior, respectively.  When $T_{\rm MI}$ is reached, an abrupt change in $S(T)$ takes place.  The inset of Fig. 2 shows the large thermal hysteresis when measurements are performed cooling down and warming up the samples.  Such a thermal hysteresis is suppressed in samples with Eu content higher than $x = 0.3$ (not shown).

\begin{figure} [htp]
\centering
\includegraphics [width=0.8\textwidth] {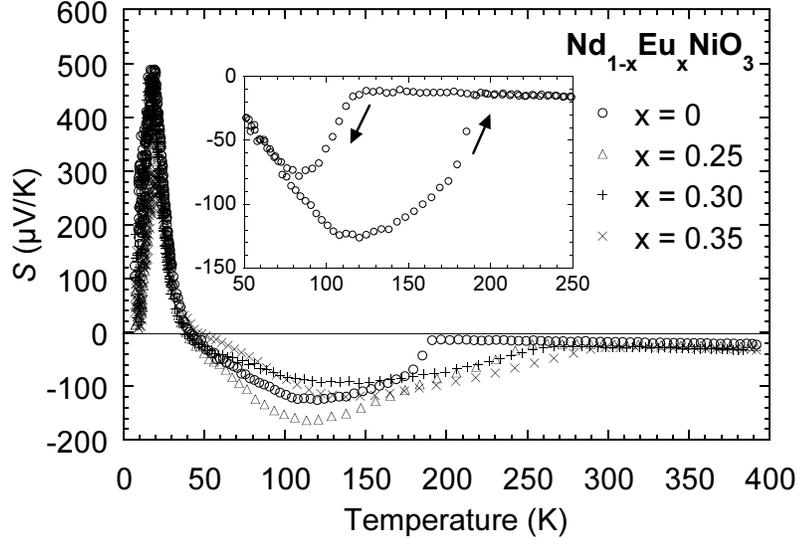}
\caption{\label{fig:epsart2} Temperature dependence of the thermopower obtained during warming the samples.  The inset displays the thermal hysteresis observed at the MI transition for the sample with $x = 0$}
\end{figure}

The low temperature behavior of $S(T)$ is much more complex and two different regions are of interest.  In the range 40 K $ \leq T \leq T_{\rm MI}$, $S(T)$ is negative, reaches $\sim -100$  V/K (for $x =$ 0), and displays a well defined minimum at a given temperature $T_{\rm m} \sim 120$ K.  Similar minimum has been observed in NdNiO$_{3}$ at $T_{\rm m} \sim 110$ K, below which $S(T)$ approached zero [16].  Such a behavior is not expected in insulators, where $S(T)$ would diverge as temperature is lowered.  Below $T_{\rm m}$, $S(T)$ decreases and becomes positive at $\sim$ 40 K.

In the range $2 < T < 40$ K, $S(T)$ is positive, indicating positive charge carriers, shows a sharp, large, and positive $\sim$ 500  V/K peak at $T_{\rm S} \sim$ 20 K.  Such a peak at $T_{\rm S}$ has been found to be Eu-independent and its shape does not follow the expected phonon-drag behavior, i. e., $T^{3}$ for temperatures below the peak and $T^{-1}$ for temperatures above.  The magnitude of $S(T)$ depends on the Eu-concentration, displaying a systematic decrease with increasing Eu content.

In order to better understand the role of the magnetic degrees of freedom in the electronic properties of these compounds, the pristine sample was subjected to an external magnetic field of $\sim 90$ kOe.  Figure 3 shows the $S(T,H)$ data for two different magnetic fields during the warming cycle.  The $S(T,H)$ data are magnetic field independent for $T > 30$ K.  In the low temperature region ($T < 30$ K), $S(T)$ is enhanced by $\sim 25 \%$ when a magnetic field of 90 kOe is applied, although the temperature where the maximum occurs is also magnetic field independent.

\begin{figure} [htp]
\centering
\includegraphics [width=0.8\textwidth] {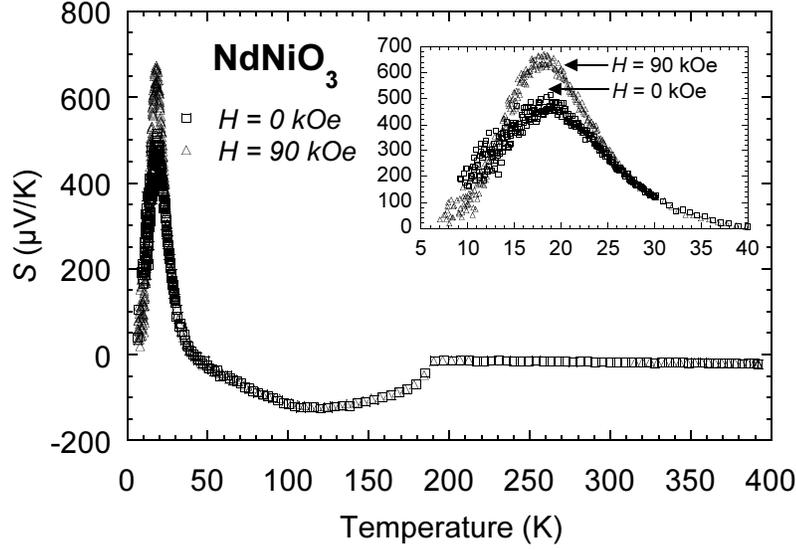}
\caption{\label{fig:epsart3} Temperature dependence of the thermopower of NdNiO$_{3}$ compounds under two applied magnetic fields.  The inset displays details of the increase of the maximum at $T_{\rm S} \sim 20$ K with increasing magnetic field}
\end{figure}

The results for $\kappa(T,H)$ (not shown) indicated that the thermal transport is not affected by the external magnetic field, further suggesting that the only one contribution to $\kappa(T,H)$ arises from the lattice.  On the other hand, the change of the sign of $S(T)$ below 50 K (see Fig. 3) suggests a competition between two types of carriers and, consequently, an inhomogeneous ground state.  Such a ground state can be visualized as composed of at least two kinds of regions (or pockets) with different charge carriers and different properties under applied magnetic fields.  As the shape of $S(T,H)$ is essentially unchanged under application of magnetic fields, one of these regions is believed to be very sensitive to the magnetic field and its thermopower properties are largely enhanced.  We also mention that the nature of the phase responsible for the large magnetothermopower observed here is certainly related to the complex magnetic behavior of the Ni sublattice below $T_{\rm N}$ and the magnetic ordering of the Nd ions at low temperatures [5-7].  In any event, neutron diffraction measurements at low temperatures are underway in order to clarify this point.

\ack This work was supported by the Brazilian agency FAPESP under Grant No. 05/53241-9.  One of the authors (R.F.J.) is a CNPq fellow under Grant No. 303272/2004-0.

\end{document}